\documentclass[prl, twocolumn,showpacs,amsmath,amssymb,superscriptaddress,nofootinbib]{revtex4-1}

\pdfoutput=1

\usepackage{graphicx}% Include figure files
\usepackage{amssymb}
\usepackage{amsmath}
\usepackage{pifont}
\newcommand{\vect}[1]{\boldsymbol{\mathbf{#1}}}
\usepackage[usenames,dvipsnames]{color}
\usepackage[colorlinks=true, linkcolor=BrickRed, citecolor=Blue, urlcolor=Blue, filecolor=Blue]{hyperref}
\usepackage[dvipsnames]{xcolor}

\newcommand{\be}{\begin{equation}}
\newcommand{\ee}{\end{equation}}
\newcommand{\bea}{\begin{eqnarray}}
\newcommand{\eea}{\end{eqnarray}}

\begin{document}

\title{Novel Spectral Features in MeV Gamma Rays from Dark Matter}

\author{Torsten Bringmann}
\email{torsten.bringmann@fys.uio.no}
\affiliation{Department of Physics, University of Oslo, Box 1048 NO-0316 Oslo, Norway}

\author{Ahmad Galea}
\email{ahmad.galea@fys.uio.no}
\affiliation{Department of Physics, University of Oslo, Box 1048 NO-0316 Oslo, Norway}

\author{Andrzej Hryczuk}
\email{a.j.hryczuk@fys.uio.no}
\affiliation{Department of Physics, University of Oslo, Box 1048 NO-0316 Oslo, Norway}

\author{Christoph Weniger}
\email{c.weniger@uva.nl}
\affiliation{GRAPPA, Institute of Physics, University of Amsterdam, Science Park
904, 1090 GL Amsterdam, Netherlands \vspace{0.3cm}}

\date{February 16, 2017}

\begin{abstract} 
\noindent 
Astrophysical searches for gamma rays are one of the main strategies to probe the 
annihilation or decay of dark matter particles. We present a new class of distinct sub-GeV spectral 
features that generically appear in kinematical situations where the available center-of-mass 
energy in such processes is just above threshold to produce excited meson states. Using a Fisher forecast
with realistic astrophysical backgrounds, we demonstrate that for upcoming experiments like e-ASTROGAM 
and ComPair these signals can turn out to be the smoking gun in the search for particle dark matter. 
\end{abstract}

%\pacs{}

\maketitle

%%%%%%%%%%%%%%%%%%%%%%%%%%%%%%%%%%%%%
\paragraph*{Introduction.---}%
Gamma rays provide a promising way of identifying the nature of dark matter (DM), not the least
because they may carry distinct spectral features that would provide a 
smoking-gun signal against dominant astrophysical backgrounds  \cite{Bringmann:2012ez}.
Those features are expected at the highest kinematically accessible energies from
DM annihilation or decay, and hence at GeV to TeV energies for DM candidates that
arise in theories extending the electroweak sector of the standard model of particle physics.
In this energy range, the most stringent limits on monochromatic, or `line', features are
presently provided by observations of the Galactic center (GC) region and 
halo \cite{Ackermann:2015lka,Abramowski:2013ax}.
At much lower energies, in the keV range, monochromatic photons may arise from the decay 
of sterile neutrinos, another excellent DM candidate \cite{Boyarsky:2009ix,Abazajian:2012ys}. 
There are, however, also nuclear
transitions that produce X-ray lines in this energy range, which must be carefully modelled
in order not to be confused with a signal (for a recent and still controversial hint of such
a signal, see Refs.~\cite{Bulbul:2014sua, Boyarsky:2014jta}). 

While these two energy bands have received a lot of attention in the context of DM searches, 
energies in the MeV range have so far been studied in much less detail -- though early work 
argued that observable 
quasi-monochromatic photons at these energies may result from DM annihilation to 
quarkonium \cite{Srednicki:1985sf,Rudaz:1986db}, as well as step-like features from the decay 
$b\to s+\gamma$ or $b'\to b+\gamma$, where
$b'$ is a hypothetical 4th generation quark  \cite{Bergstrom:1988jt}. Another possibility is
the decay of DM candidates like the gravitino, which has motivated a dedicated line search 
with the Fermi Large Area Telescope down to energies of 100 MeV \cite{Albert:2014hwa}. 
It was also pointed out
that for DM lighter than around 100 MeV, the only kinematically accessible non-leptonic states 
are photons and neutral pions, leading to clear gamma-ray signatures to look for  
\cite{Boddy:2015efa,Boddy:2015fsa,Boddy:2016fds}. At those energies, however, there is 
a significant `MeV gap' \cite{Greiner:2011ih} in the sensitivity of operating and past 
experiments, such that presently only very weak limits on DM signals exist in this range \cite{Essig:2013goa}.

There is already a strong interest in the astrophysics community to finally fill this MeV gap, via 
planned missions like \mbox{e-ASTROGAM} \cite{e-ASTROGAM} and 
ComPair \cite{Moiseev:2015lva}, in order to 
address a broad key science program ranging from the physics of ultra-relativistic jets to 
a better understanding of the Galactic chemical evolution. Here, we point 
out a new class of potential smoking-gun signatures for DM signals in the range  
10\,MeV$\lesssim E_\gamma\lesssim$\,100 MeV, providing further motivation for the realization of 
such missions. These signatures involve transitions between meson states and, in their 
simplest realization,  do not require any new physics (beyond, obviously, the DM 
particle itself) but inevitably arise in certain kinematical situations for GeV-scale DM annihilating or 
decaying to heavy quarks. Unlike direct detection or collider experiments, these signatures
are thus very sensitive to DM coupling with third or second generation quarks.

This article is organized as follows. We first briefly review the standard arguments for 
a featureless gamma-ray spectrum from DM, and then illustrate for the case of 
$B$ and $D$ mesons how the production and decay of excited meson states can change the 
picture at sub-GeV photon energies. 
We then adopt the characteristics of planned experiments in the MeV range
for a detailed Fisher forecast, demonstrating
that the spectral features identified here can significantly help to discriminate DM signals
from astrophysical backgrounds. We move on to discuss further expected features in this energy
range and then present our conclusions, along with an outlook for future directions
of investigation. In two Appendices we assess the impact of the assumed 
experimental settings, and provide details about the adopted Fisher forecast.

\begin{figure*}[t!]
\includegraphics[width=0.6\columnwidth]{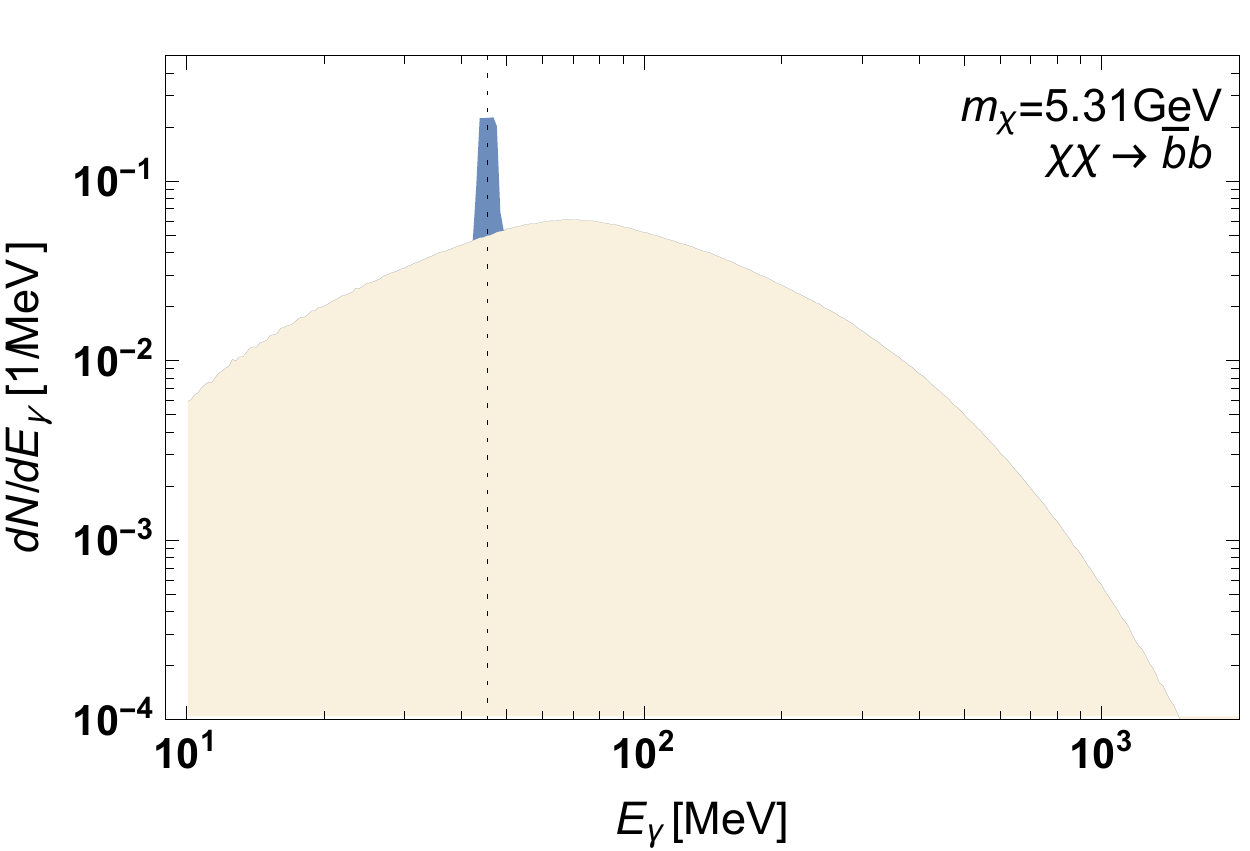}~~~
\includegraphics[width=0.6\columnwidth]{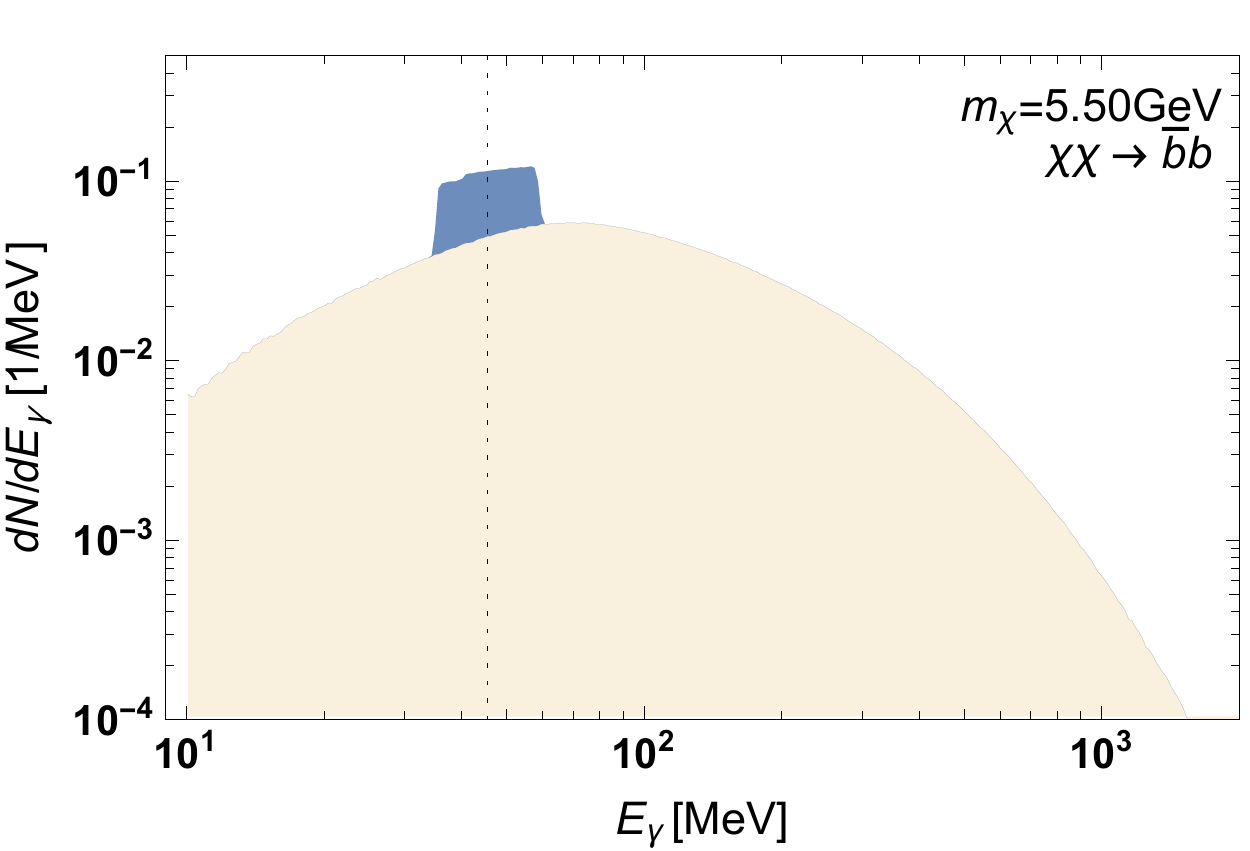}~~~
\includegraphics[width=0.6\columnwidth]{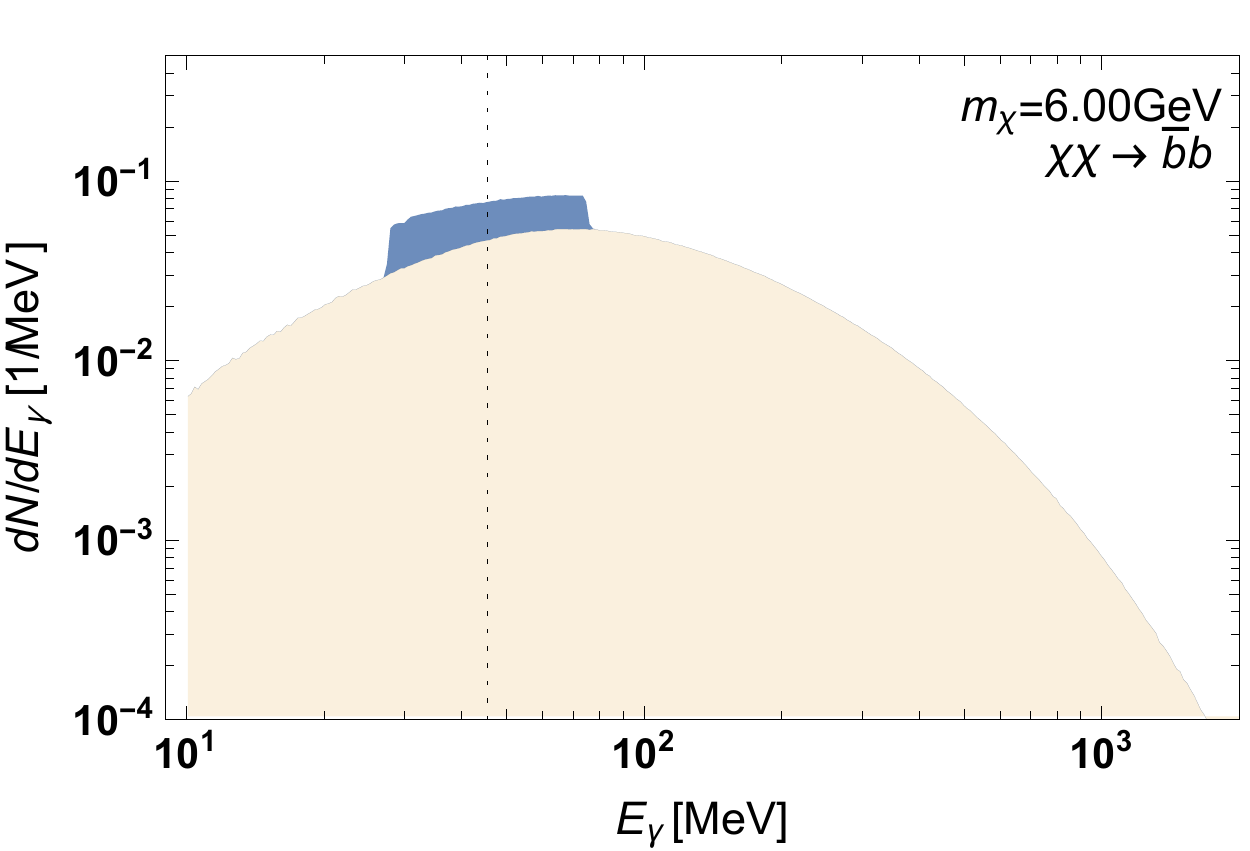}
\includegraphics[width=0.6\columnwidth]{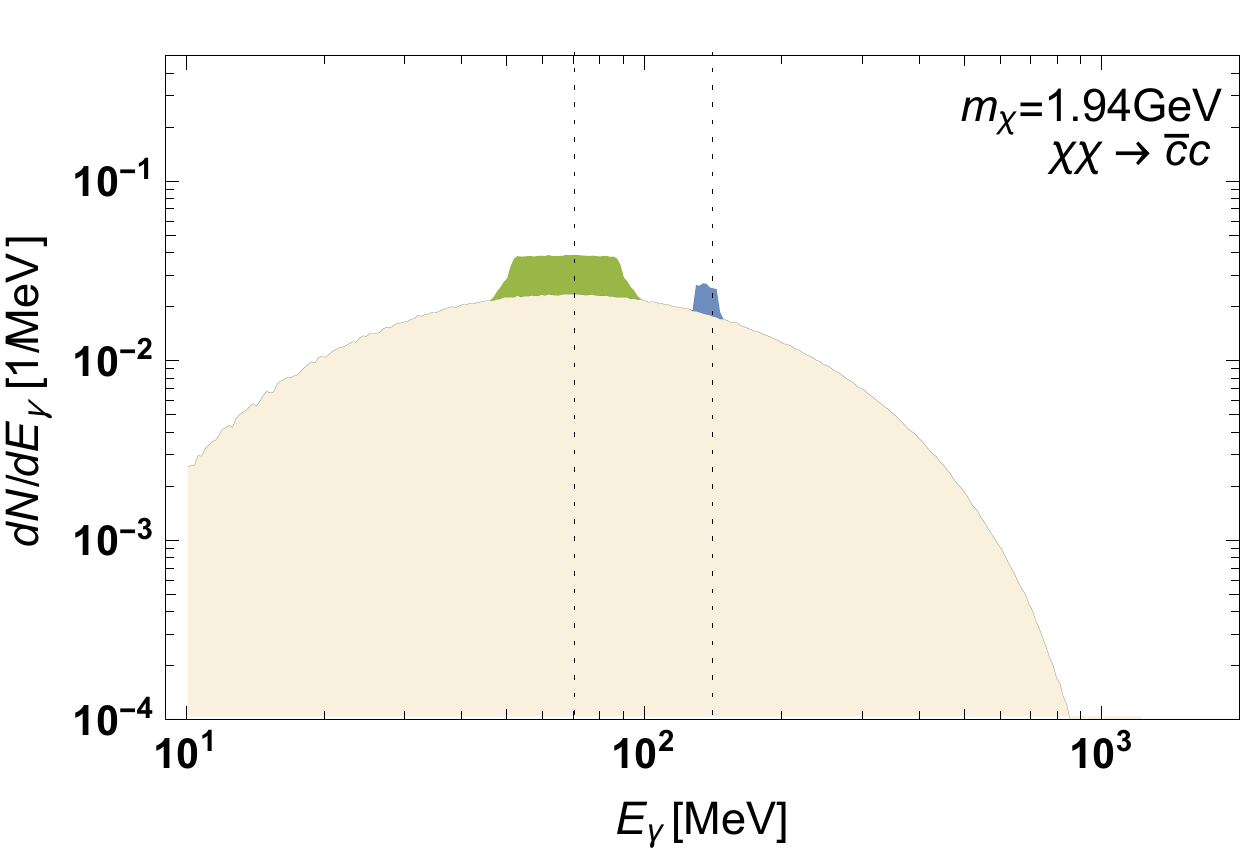}~~~
\includegraphics[width=0.6\columnwidth]{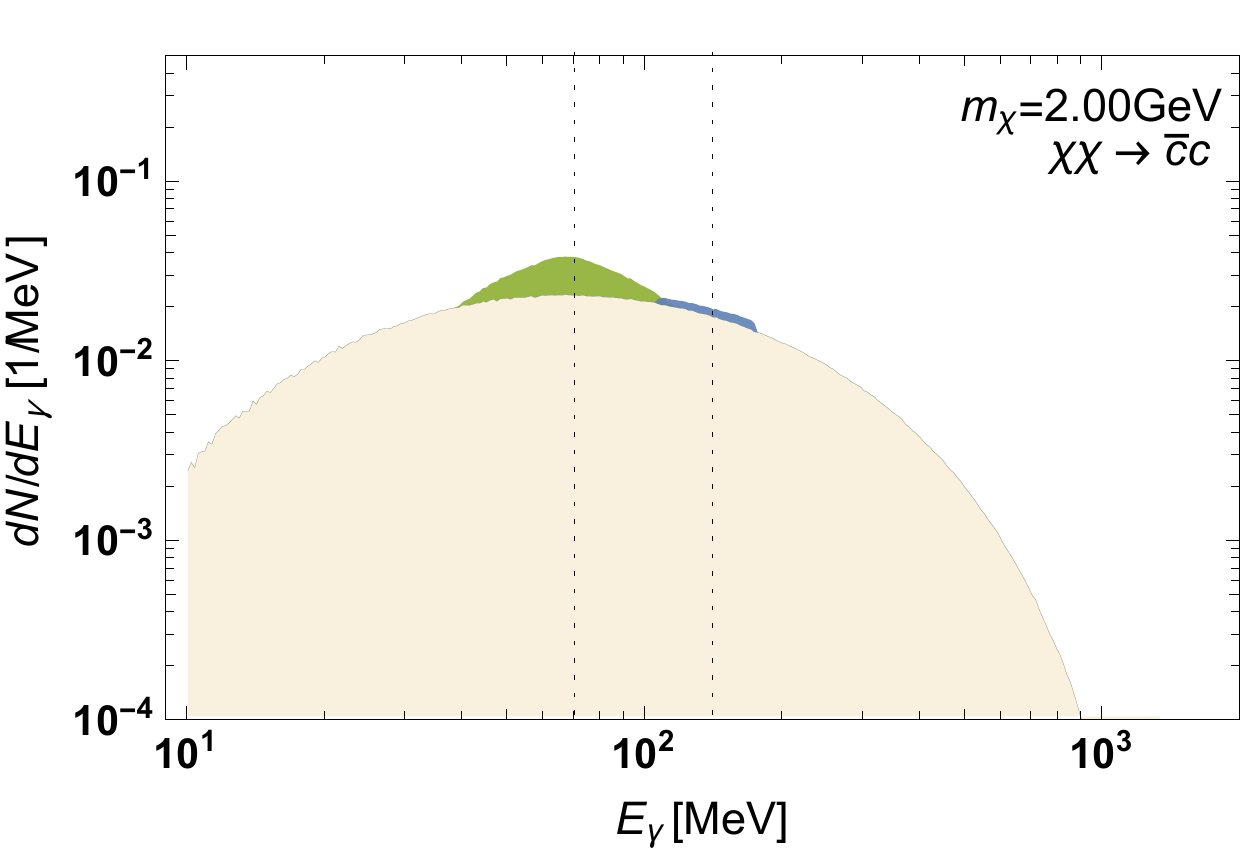}~~~
\includegraphics[width=0.6\columnwidth]{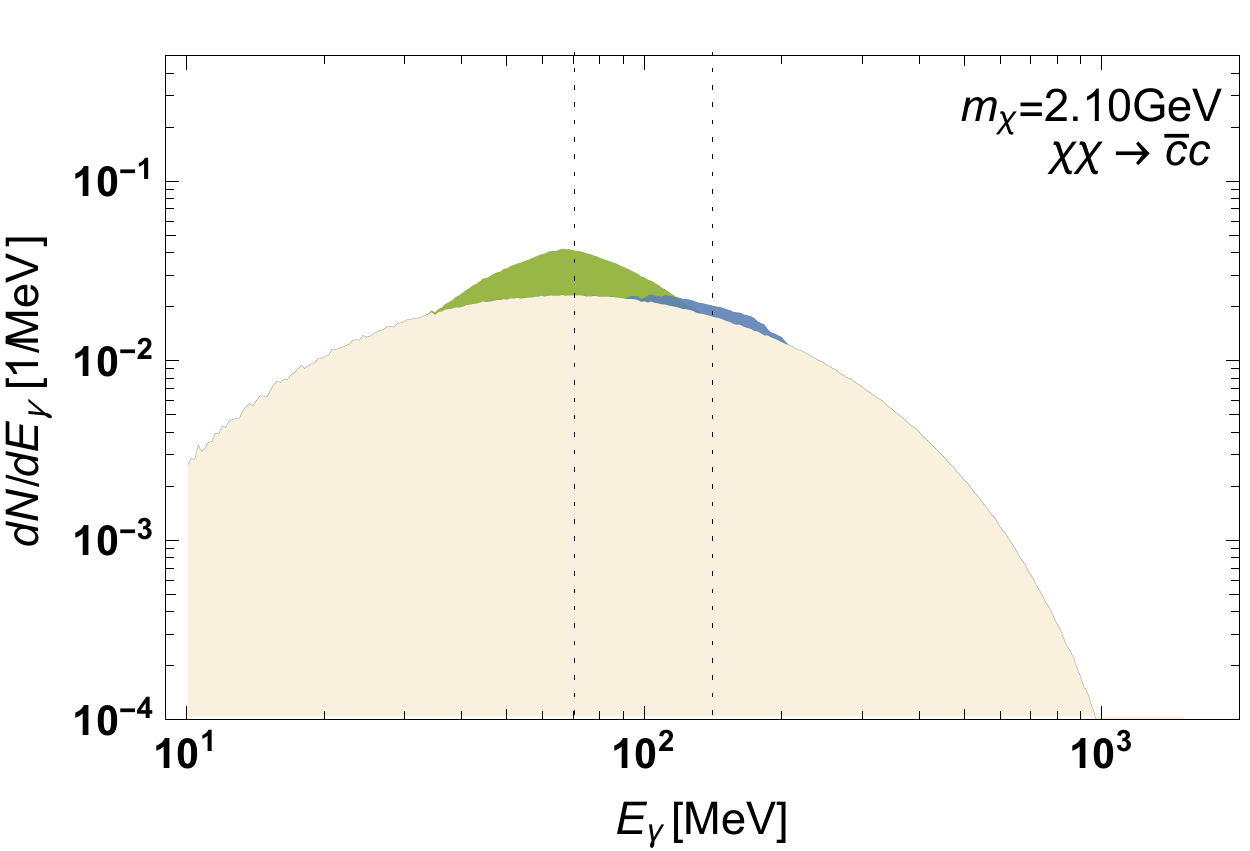}
\caption{(\emph{Top row}) 
Gamma-ray spectra from $\chi\chi\to\bar bb$, for increasing DM 
mass $m_\chi$ (from left to right). The light shaded part is the standard continuum
contribution, dominated by $\pi^0$ decay. The dark shaded feature results from the decay
of excited $B$ meson states, $B^*\to B+\gamma$;  the vertical dashed line indicates 
the corresponding average mass difference $\Delta M_B\equiv 0.046$\,GeV.
(\emph{Bottom row}) Same, but for $\bar c c$ final states. The {\it two} 
pronounced features here arise from excited $D$ mesons, namely  
$D^*\to D+\pi^0$, $\pi^0\to\gamma\gamma$ (left) and $D^*\to D+\gamma$ (right).
The latter is roughly centered on $\Delta M_D\equiv 0.142$\,GeV, the former on $\Delta M_D/2$.
}
\label{fig:spectra}
\end{figure*}

%%%%%%%%%%%%%%%%%%%%%%%%%%%%%%%%%%%%%
%%%%%%%%%%%%%%%%%%%%%%%%%%%%%%%%%%%%%
\smallskip
\paragraph*{Meson spectroscopy with dark matter.}%
The annihilation or decay of DM typically 
produces, through decay and fragmentation
of the final state particles, a large number of neutral pions with energies all the way up to 
what is kinematically accessible in a given process. Those pions decay dominantly via
$\pi^0\to\gamma\gamma$, resulting in two monochromatic photons with 
$E_\gamma=m_{\pi^0}/2$ in the respective pion's rest frame. When boosted to the DM frame, 
taking into account the high multiplicity of the pions, this leads to 
a featureless gamma-ray spectrum that is almost indistinguishable among all
quark and weak gauge boson final states \cite{Bringmann:2012ez}.

In this article we point out that there are interesting exceptions to this simple, yet widely spread 
picture. In fact, this should not come as a surprise in view of the highly complicated 
multi-step decay and fragmentation cascades that {\it actually} take place in a given annihilation
or decay process, and which must be simulated with event generators like {\sf Pythia} 
\cite{Sjostrand:2006za,Sjostrand:2014zea} or {\sf Herwig} \cite{Corcella:2000bw} to arrive at 
general conclusions like the one just quoted. 
%In particular, pions are by no means the only hadronic states produced in these events.
Concretely, heavier mesons and baryons are formed as soon as allowed by kinematics,  with the 
former, requiring only two quarks to combine, being much more abundant than the 
latter.
In the ground state, heavy mesons mostly decay directly to lighter mesons and 
leptons \cite{Agashe:2014kda}, leading to cascades that eventually result in pions.
Large mass hierarchies, furthermore, generally imply that intermediate states in such
showering process are produced with high virtuality, which in turn leads to a large probability of 
gluon emission and therefore again high multiplicities of lighter 
states \cite{Gribov:1972ri,Altarelli:1977zs,Dokshitzer:1977sg}. % [DGLAP papers]. 

If, on the other hand, a meson containing heavy quarks is produced in an excited state, it will 
typically de-excite before decaying to a lighter meson type with a different quark content 
-- most often by emitting 
a monochromatic photon or pion.  In both situations a clear spectral features arises 
in the DM rest frame: due to the non-zero kinetic 
energy of the excited meson the monochromatic photon leads to a box-shaped 
spectrum roughly centered on the energy difference between the meson states:%\footnote{%
% More correctly, the spectrum is given by 
\be
\frac{dN}{dE_\gamma}=\frac1{E_\mathrm{max}-E_\mathrm{min}}\theta(E_\gamma-E_\mathrm{min})
 \,\theta(E_\mathrm{max}-E_\gamma),
\ee
where $\theta$ is the Heaviside function and 
$E_\mathrm{max,min}=E_\gamma'(E^*/M^*)\left(1\pm\beta \right)$.
Here, $E^*$ and $M^*$ are the energy and mass of the excited meson (in the DM frame)
and $\beta=(1-{M^*}^2/{E^*}^2)^{1/2}$ its velocity; $E_\gamma'=\Delta M\left( 1-\Delta M/(2M^*)\right)$  
is the photon energy in the  decaying meson frame and $\Delta M$ the mass difference to the ground state. 
Photons from $\pi^0\to\gamma\gamma$, on the other hand, give a bump centered on half of this energy.
Both features become wider with larger kinetic energy of the initial meson; in 
practice, they are sufficiently pronounced only in situations where 
the excited meson
is nearly at rest. In this situation, the location and shape of the resulting spectral features 
in gamma rays does not only allow for an accurate determination of the DM mass, but in principle 
also provides a  direct way of inferring both the initial meson state and 
the de-excitation channel.

%%%%%%%%%%%%%%%%%%%%%%%%%%%%%%%%%%%%%
\smallskip
\paragraph*{Example spectra.}%
In order to illustrate these considerations, let us now concentrate on DM annihilation to 
$\bar bb$ or $\bar cc$. In this case virtually every resulting shower will contain at least 
one $B$ or $D$ meson, respectively, from the hadronization of one of the final state particles
with a light quark. The relevant excited $B$ meson states are mostly $B^{-*}$ and $B_0^{*}$, 
which decay by emitting a photon with energy 
$m_\gamma=m_{B^{-*}}-m_{B^{-}}= 0.046 \,\rm{GeV}$ 
and $m_\gamma=m_{B_0^{*}}-m_{B_0}= 0.045 \,\rm{GeV} $ \cite{Agashe:2014kda}. The 
mesons ($D^{-*},D_0^{*},D_s^{-*}$), on the other hand, decay via {\it both} decay channels discussed above 
to the respective ground state; this produces neutral pions and photons with an 
energy of ($0.140, 0.142, 0.144$)\,GeV and branching ratios of $BR_{D^*\to D\pi^0}\approx2/3$  
and $BR_{D^*\to D\gamma}\approx1/3$ \cite{Agashe:2014kda}.

In Fig.~\ref{fig:spectra}, we show the resulting photon spectra for DM annihilation into 
$\bar bb$ and $\bar cc$, for a number of benchmark values for the DM mass (the same spectra 
arise for the {\it decay} of a DM particle with twice the stated mass). In order to produce these 
plots, we ran {\sf Pythia} v8.215 \cite{Sjostrand:2007gs} to simulate $10^6$  events with an initial state 
back-to-back $\bar qq$ pair and a center-of-mass energy of $2\,m_\mathrm{DM}$, adopting
default tuning settings and including both photon and gluon final state radiation. The expected
box features around $E_\gamma\simeq\Delta m$ from monochromatic photons are clearly visible, 
as well as -- for the case of $\bar cc$ final states -- a second feature around 
$E_\gamma\simeq\Delta m/2$ from monochromatic neutral pions. These features appear on top
of the dominant contribution from photons that results from $\pi^0$ produced at all energies in 
the fragmentation process. Increasing the DM mass, the new spectral features that we have 
reported here broaden and relatively quickly become indistinguishable from the the standard pion bump.

\begin{table}[t!]
 \centering
 \begin{tabular}{lrrrrr}
   \hline
   \hline
   Experiment & $\Delta E/E$ & FoV [sr] & $A_\textrm{eff}$ [$\rm{cm}^2$] &
   $T_{\rm obs}$ \\
   \hline
   e-ASTROGAM  \cite{e-ASTROGAM, Knodlseder:2016pey}
   & 25\% & 2.5 & 1500 & 5 yr\\
   ComPair  \cite{Moiseev:2015lva}
   & 12\% & 3 & 1000 & 5 yr\\
   \hline
   \hline
 \end{tabular}
 \caption{Adopted characteristics for upcoming or planned instrument in the
   pair-production regime for the $10$–-$3000\,\rm{MeV}$ energy range
   ($10$--$1000\,\rm{MeV}$ in the case of ComPair).  For the
   energy resolution we adopt the value at 100 MeV close to the spectral
   features of interest.  For the adopted ROI the finite angular resolution is
   irrelevant.  We will assume survey mode with equal sky coverage throughout.
   }
 \label{tab:instruments}
\end{table}

%%%%%%%%%%%%%%%%%%%%%%%%%%%%%%%%%%%%%
%%%%%%%%%%%%%%%%%%%%%%%%%%%%%%%%%%%%%
\smallskip
\paragraph*{Detecting features in the MeV gap.}
There is a pronounced interest of the gamma-ray astronomy 
community to improve the coverage of sub-GeV photon energies. During the last years,
this has culminated in two active efforts for medium-sized satellite
missions.  Firstly e-ASTROGAM~\cite{e-ASTROGAM}, which is 
%carried mostly by the European community and 
proposed as an ESA M5 mission by the European community; and secondly
ComPair~\cite{Moiseev:2015lva}, which is a proposal mostly carried by the US
community.
In order to assess the expected detection significance of the spectral features described 
above, we will in the following adopt the preliminary characteristics of these detectors 
as summarized in Tab.~\ref{tab:instruments} and perform a Fisher forecast.  
The Fisher forecast takes into account the  full covariance matrix of the spectral analysis, 
for which convenient analytical expressions are presented in Ref.~\cite{Fisher}.

We model the differential flux by $\phi(E, \vec\theta) = \phi_{\rm sig} + \phi_{\rm bg}$, 
where $\vec\theta$ are the model parameters.  The model is assumed to be linear
in $\vec\theta$.  The Fisher information matrix for a
spectral analysis and parameters $\theta_i$ and $\theta_j$ is then given
by (see Appendix B
and Ref.~\cite{Fisher})
\begin{equation}
  \mathcal{I}_{ij} = T_\text{obs} A_\text{eff} \int_{E_\text{min}}^{E_\text{max}}
  dE \frac{\partial_i\phi(E)
  \partial_j\phi(E)}{\phi_\text{bg}(E)} + \delta_{ij}\frac{1}{\Sigma_i^2}\;,
  \label{eqn:fish}
\end{equation}
where $T_\text{obs}$ is the observation time, $\partial_i\phi(E)$ denotes the
change in the differential flux as function of parameter $\theta_i$, and
$\phi_\text{bg}(E)$ is the expected observed flux (assumed to be dominated by
  the background).  As energy range, we always adopt $E_\text{min}, E_\text{max}
= 10{\rm\,MeV}, 1{\rm\,GeV}$, to allow for an easy comparison between
instruments. Lastly $\Sigma_i^2$ refers to the external variance of parameter
$\theta_i$, e.g.~from additional external knowledge of the background
systematics.

\begin{figure}
  \includegraphics[width=0.85 \columnwidth]{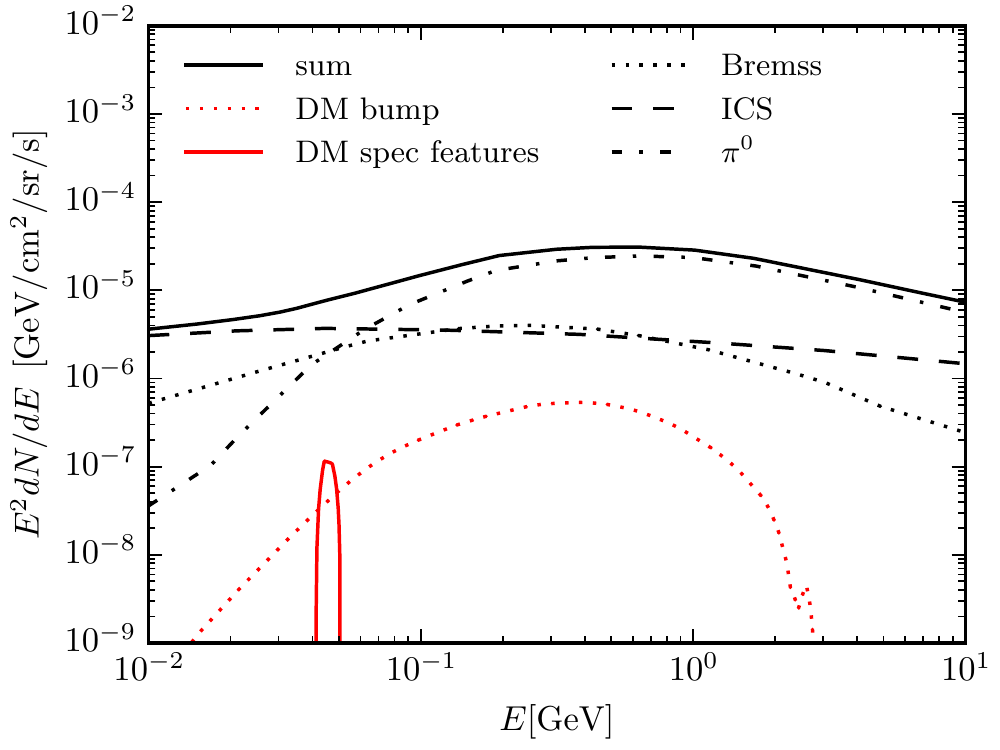}
  \caption{Backgrounds \cite{Strong:2011pa} and one variant of the signal spectra used in the current
    analysis ($\chi\chi\to\bar bb$ with $m_\chi=5.3$\,GeV).
    }
    \label{fig:BGspectra}
\end{figure}

We model the background with the three components shown in
Fig.~\ref{fig:BGspectra}, taken from Ref.~\cite{Strong:2011pa}.  Our region of interest (ROI) 
is a $20^\circ\times 20^\circ$ region around the GC;  we approximate
its intensity by the more extended ROI used in \cite{Strong:2011pa}. 
%, which is more extended along the Galactic disk. 
In the
Fisher analysis, we allow not only the normalization of each of the three
components to vary, but also their slopes and the curvatures.  Hence, our
complete background model reads
$\phi_{\rm bg}=\sum_{i=1}^3(\theta_i^n+\theta_i^s \log(E/E_0) +
\theta_i^c\log(E/E_0)^2)\phi_i$, where $i=1,2,3$ refers respectively to inverse 
Compton scattering (ICS),
bremsstrahlung and the astrophysical $\pi^0$ contribution; $E_0 = 0.3\rm\,GeV$ is a pivot point, 
and ($\theta_i^n=1$, $\theta_i^s = \theta_i^c = 0$) describe the
baseline model.  For the external variance of the background parameters, we
assume standard deviations $\Sigma_{\theta_i^n}=0.5$,
$\Sigma_{\theta_i^s}=0.15$ and $\Sigma_{\theta_i^c}=0.05$.  These numbers imply
that, within $2\sigma$ variance, the background model can vary by roughly a
factor two in the considered energy range.  This is adopted \textit{ad hoc},
and more accurate estimates can only be made once data is available.  Our
qualitative conclusions are relatively insensitive to this number.

The signal is modeled by two components as also shown in Fig.~\ref{fig:BGspectra},
$\phi_{\rm sig}=\sum_{i=4}^5 \theta_j \phi_i$, where $i=4,5$ corresponds
respectively to the broad pion bump and the spectral features visible in Fig.~\ref{fig:spectra}.  
In the case of $\bar cc$ final states, we treat the two spectral features together.  
For the sake of this figure, the
spectra are normalized to a reference cross-section of $\langle\sigma
v\rangle=10^{-26}\rm\,cm^3\,s^{-1}$.  We adopt a standard Navarro-Frenk-White (NFW) 
profile with a
scale radius 20~kpc, $0.4\rm\,GeV\,cm^{-3}$ local density and 8.5~kpc distance to
the GC (see Ref.~\cite{Bringmann:2012ez} for details).  The corresponding $J$-value
integrated over the ROI is $5.0\times10^{22}\,\rm GeV^2\,cm^{-5}$. 
%Only the normalization is left as unconstrained parameter.
% (which formally corresponds to $\Sigma_{4,5}\to\infty$).

In total we are thus dealing with an eleven parameter model.  The expected variance
of the DM signal normalization parameters is
$\sigma^2_{ii}=(\mathcal{I}^{-1})_{ii}$, with $i=4,5$.  Here,
$\mathcal{I}^{-1}$ denotes the inverse of the $11\times11$ Fisher information
matrix.  Note that the matrix inversion fully takes into account correlations
between background and signal components in the model.  The projected
$95\%\rm\,CL$ upper limit on the annihilation cross section into spectral
features is then $\langle\sigma v\rangle_{\rm UL} = 1.65\cdot\sigma_{55}\cdot
10^{-26}\rm\,cm^3\,s^{-1}$ (see Appendix B). %Appendix~\ref{apx:fisher}).

\begin{figure}
  \includegraphics[width=0.85 \columnwidth]{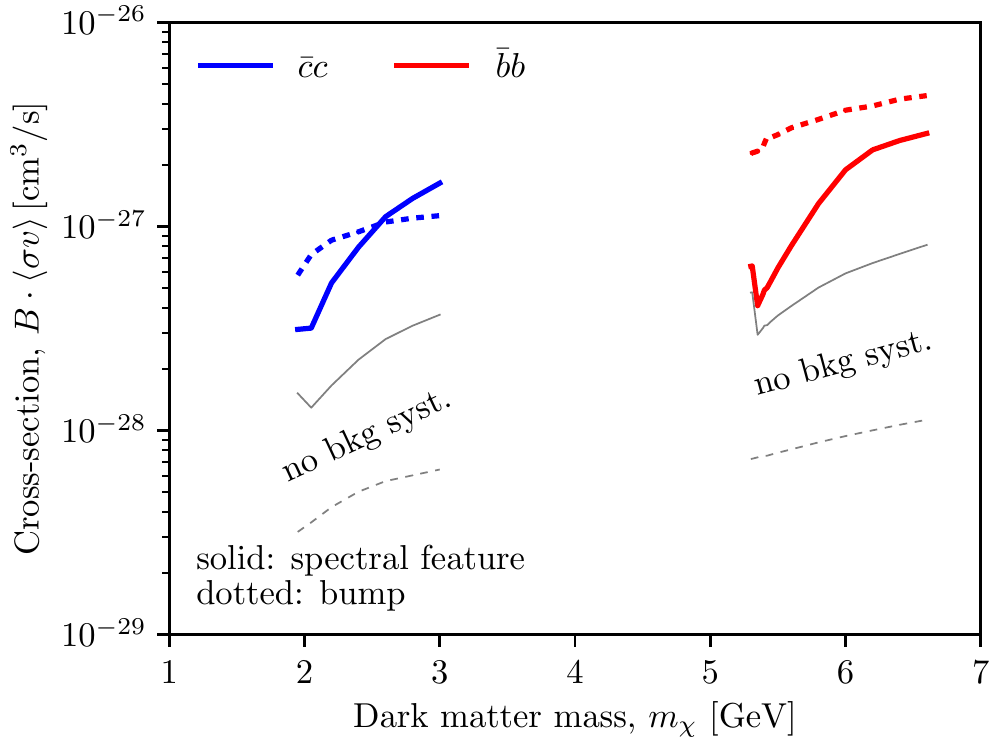}
  \caption{The solid and dotted red and blue lines show projected $95\%$\,CL upper limits
    on the spectral features as well as the pion bump, including our estimate
    for the background systematics, for the ComPair satellite.  We also show
    (in gray) results obtained when neglecting background systematics. For
    profiles steeper than NFW, and an optimized ROI, $B\sim10$ is possible (see text for 
    discussion).}
  \label{fig:UL}
\end{figure}

%%%%%%%%%%%%%%%%%%%%%%%%%%%%%%%%%%%%%
%%%%%%%%%%%%%%%%%%%%%%%%%%%%%%%%%%%%%
\smallskip
\paragraph*{Results.} Our results for the projected upper limits are summarized in
Fig.~\ref{fig:UL}. Here, we consider for illustration only ComPair; see
the supplemental material for similar results for e-ASTROGAM.  
We show the projected $2\sigma$ upper limits that
could be obtained for DM masses close to the kinematic cutoff for the
indicated quark channels. 

We find that, indeed, after taking into account a realistic
model for background uncertainties, the spectral features (solid lines) have a larger
constraining power than the broad pion bump (dotted lines).  If one were to completely 
neglect background systematics (light gray lines), one would falsely conclude that 
the pion bump is more constraining. Note that at very small masses the limits on the spectral 
features become slightly less constraining again; this is because {\it some} of the excited meson 
states are no longer kinematically accessible.

%%%%%%%%%%%%%%%%%%%%%%%%%%%%%%%%%%%%%
%%%%%%%%%%%%%%%%%%%%%%%%%%%%%%%%%%%%%
\smallskip
\paragraph*{Discussion.}

While our projected limits from the pion bump alone would 
already be competitive with present bounds from dwarf galaxy observations
by the Fermi gamma-ray space telescope \cite{Ackermann:2015zua}, including 
the spectral features in the analysis would significantly improve them. Let us
stress, however, that Fig.~\ref{fig:UL} mainly serves to illustrate the {\it relative} 
importance of the two signal contributions in setting the limit. Rather than the 
annihilation rate $\langle\sigma v\rangle$ we hence plot $B \langle\sigma v\rangle$, 
where $B=1$ corresponds to the specific analysis settings described above. Both 
a data-optimized ROI (see e.g.~\cite{Bringmann:2012vr}) and a DM profile steeper 
than NFW would easily increase $B$ by a factor of a few, allowing ComPair or 
e-ASTROGAM to detect the spectral features described here even if there is no hint 
for a signal in dwarf galaxy observations.

Concerning possible spectral features, the $B$ and $D$ meson families we have focussed
on here have the advantage of de-exciting via the emission of a single photon or neutral 
pion. Furthermore, while DM annihilation or decay can directly produce such excited states with
small kinetic energies, this is not expected for astrophysical processes. 
Let us stress,
however, that the spectra shown in Fig.~\ref{fig:spectra} are just {\it examples} for
similar features that may arise at sub-GeV energies.
%In general, desirable features include: as large photon energy relative to the decaying 
% meson mass as possible, several transition lines close in energy (i.e. existence of more 
% than one excited state that gives similar transition).
%Direct production implies that we are intersted in mesons with at least one light quark: one of the 
% quarks need to be produced out of breaking of the gluonic flux tubes and the probability fall
% down exponentially with mass of the quark.

The dark sector may, {\it e.g.}, feature a non-Abelian gauge symmetry with confinement~\cite{Kribs:2016cew}. 
The dominant final states of DM annihilation would then naturally be dark meson states that de-excite by 
emitting a dark pion $\tilde \pi$.
%, in full analogy to the case of standard model mesons. 
If $\tilde\pi$ dominantly decays to two photons, 
%even if very long-lived, 
this would lead to identical 
features as for the decay of a standard $\pi^0$ -- with the difference that these features 
could in principle appear at  {\it any} energy because the differences in energy 
levels follow from the physics of the dark and not the visible sector.
As noted earlier \cite{Srednicki:1985sf,Rudaz:1986db,Bergstrom:1988jt}, 
DM annihilation to bound quark-antiquark states also leads to potential smoking-gun signatures 
if accompanied by the emission of a (necessarily quasi-monochromatic) photon.
%quasi-monochromatic because of the finite quarkonium lifetime.  
We therefore expect {\it further} identifiable features if the quarkonium is not 
produced in its ground state or if the co-produced boson is a $\pi^0$ rather than a photon. 
While this adds yet another promising type of sub-GeV spectral features to our list, 
a full classification of the potentially rich phenomenology is beyond the scope of the present work.

Let us finally stress that codes like {\sf Pythia} are tuned to higher energies, where the formation 
of the $\bar q q$ pair and the subsequent hadronization can be treated 
as separate processes. This clearly introduces a certain theoretical error,
warranting more detailed studies about meson production at threshold (as well as direct 
quarkonium production, see the discussion above, which is not covered by {\sf Pythia}). On the other 
hand, we note that spectral features like the ones shown in Fig.~\ref{fig:spectra} arise mainly due to 
kinematics, because only a few meson states are kinematically accessible and the de-excitation 
time scale is shorter than the decay time scale. For that reason,  we do not expect that an improved 
estimate of the dynamics of meson production will lead to qualitative differences in the 
{\it relative} normalization of the spectral components.

%%%%%%%%%%%%%%%%%%%%%%%%%%%%%%%%%%%%%
%%%%%%%%%%%%%%%%%%%%%%%%%%%%%%%%%%%%%
\smallskip
\paragraph*{Conclusions and Outlook.}
The clear identification of a DM signal above astrophysical backgrounds 
generally proves to be a big challenge, and finding distinct
%line- or box-like 
spectral features on top of an observed smooth excess could be central
to such an endeavour. 
%unambiguously identify a DM signal, even if the formal statistical
%significance of the features would be comparably low. 
In this article, we have pointed out a potentially large class of such spectral 
features in the almost unexplored sub-GeV energy range.
By means of a Fisher forecast,  which in the way it is implemented here
introduces a new method in the context of indirect DM searches~\cite{Fisher},
we verified that missions like ComPair and e-ASTROGAM could indeed 
sufficiently reduce the astrophysical background uncertainties to identify such a smoking
gun signature for GeV particle DM.

We note that the possibility to probe light DM is also interesting because of the 
strongly limited sensitivity of direct detection experiments in this mass range \cite{Cushman:2013zza}
(though there are various ideas to overcome these difficulties, 
e.g.~\cite{Essig:2011nj, Essig:2012yx, Essig:2015cda, Hochberg:2015pha, Profumo:2015oya}). The 
features reported here have, furthermore, the potential to directly probe -- and in fact
disentangle -- DM couplings to 2nd or 3rd generation quarks, for which both collider 
and direct DM searches are generally less sensitive. Let us finally stress that mesons do
not only decay via photons and neutral pions; this may lead to corresponding spectral features
also in other indirect detection channels, notably positrons and neutrinos.
Taken together, this points to a potentially rich DM phenomenology at sub-GeV energies
which will open promising avenues for future studies.
% and clearly warrants a more thorough
%theoretical treatment of the underlying meson production and decay in these processes.

\vfill
%%%%%%%%%%%%%%%%%%%%%%%%%%%%%%%%%%%%%
\paragraph*{Acknowledgments.---}
We thank Richard Bartels, Lars Bergstr\"om, Lars Dal, Aldo Morselli, Julie
McEnery and Are Raklev for very fruitful discussions.
AH is supported by the University of Oslo through the Strategic Dark Matter 
Initiative (SDI).  CW is supported by the Netherlands Organization for
Scientific Research (NWO) through a Vidi grant.

%\newpage \mbox{ }
\newpage

\appendix
%%%%%%%%%%%%%%%%%%%%%%%%%%%%%%%%%%%%%

\section*{A. Experimental sensitivity to MeV features}

In the main text, we have explicitly shown the projected experimental sensitivity only for the ComPair
satellite, with adopted experimental characteristics as summarized in Tab.~1. %\ref{tab:instruments}.
Here, we will complement this by discussing the analogue to Fig.~3 %\ref{fig:UL} 
also for the e-ASTROGAM
mission and a fiducial future experiment with even better performance.

\subsection{e-ASTROGAM}
\begin{figure}[t!]
  \includegraphics[width=\columnwidth]{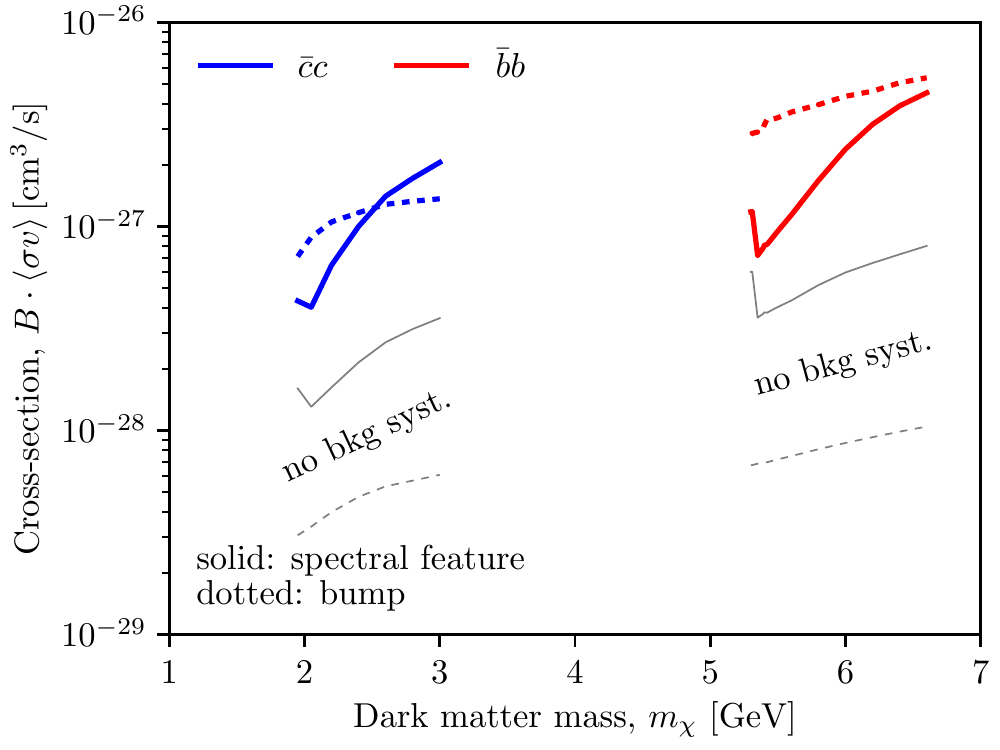}
  \caption{Same as Fig.~3 %\ref{fig:UL} 
  but for experimental characteristics corresponding to e-ASTROGAM.
  \label{fig:UL_AG}}
\end{figure}

In Fig.~\ref{fig:UL_AG}, we show the projected upper limits for e-ASTROGAM,
assuming experimental characteristics as summarized in
Tab.~1. %\ref{tab:instruments}. 
For a naive analysis, which does not include the
effect of background systematics, these limits are essentially identical to
those of ComPair (to within $10\%$, except for the close-to-threshold limits
for the $\bar bb$ channel where the difference is slightly larger).  This is
expected because the grasps of the two instruments are very similar.

Once we include the background systematics, however, ComPair is clearly
somewhat better suited to distinguish DM signal features at MeV energies than
e-ASTROGAM. This is because it has an energy resolution that is almost twice
as good, which helps to identify both the broad and the narrow spectral feature
in the DM signal. However, given that the relevant
spectral features are not more narrow than about $10\%$ for most of the parameter 
space shown in the figure, which is comparable to
the energy resolution of ComPair, the difference in general remains small -- except
for the line-like feature in the $\bar b b$ final state for $m_\chi\lesssim 5.5$\,GeV,
where ComPair becomes more sensitive by up to a factor of 2.

We finally note that for both, ComPair and e-ASTROGAM, we use the energy range
10 MeV -- 1 GeV in our analyses.  This allows an easy comparison of the
results.  However, we find that the projects constraints on the continuum
component of the DM signal significantly depend on the high-energy cutoff.  If
we extend the energy range up to 3 GeV, the e-ASTROGAM limits strengthen by a
factor of up to three for $\bar cc$ final states, and by less than two for
$\bar bb$ final states.  The projected limits for the line component change
only very mildly.

\subsection{Idealized gamma-ray experiment}

Let us now assess by how much the situation could be improved for an idealized
future experiment, for which we assume an effective area of
$A_\mathrm{eff}=10^4$\,cm$^2$ (again for an exposure of 5 years) and an energy
resolution of 1\%.  The resulting projected upper limits are shown in
Fig.~\ref{fig:UL_toy}.

As expected, the limits excluding the effect of background systematics simply improve
by a factor of roughly 3, compared to our projections for ComPair, corresponding to the 
square root of the increase in exposure.
It also becomes clear that increasing the energy resolution beyond 10\% has no 
impact on the continuum limits, even when taking into account background systematics. 
For the spectral features we are interested in here, however, a better energy resolution
would indeed imply even better detectional prospects. Taken at face value, this would
allow to constrain the $\bar bb$ channel for DM annihilation just above threshold by
almost two orders of magnitudes more stringently than current limits \cite{Ackermann:2015zua}.

\section*{B. Fisher forecast}
\label{apx:fisher}

Fisher forecasting is a common method for experimental design, and extensively
used in \textit{e.g.}~the cosmology community~\cite{Wolz:2012sr,
Khedekar:2012sh, Sellentin:2014zta}.  It is based on the
\textit{Fisher information matrix}, which is a measure of the information that
an observation is expected to carry about a set of unknown parameters.
However, its use in the indirect and direct DM detection communities is up to
now rather limited (see \textit{e.g.}~Ref.~\cite{Camera:2014rja} for previous examples).
Here, and to the best of our knowledge for the first time, we adopt some new
and simple expression for the calculation of the Fisher information matrix that
can be used for predicting sensitivities of any counting experiment in the large-number limit.  

Let us first briefly summarize the derivation of Eq.~(2), %\eqref{eqn:fish}, 
before we 
illustrate how to translate projected limits to detection sensitivities in the particular 
case we are interested in here.
The full details and a few examples will be presented elsewhere~\cite{Fisher}.  The starting
point is the unbinned Poisson likelihood function
\begin{equation}
  \mathcal{L}(\vect \theta | \mathcal{D}) =
  e^{-\mu_\text{tot}(\vect\theta)}\prod_{i=1}^{n_\text{ev}}\Phi_\text{tot}(E_i
  |\vect\theta)\;,
\end{equation}
where $\bf\theta$ denotes the model parameters, $\mu_{\rm tot}$ the total
predicted number of events, $i = 1, \dots, n_\text{ev}$ runs over the number of
measured photons, $\Phi_\text{tot}(E_i|\bf\theta)$ is the differential number
of expected photons, and $E_i$ is the energy of photon $i$.  Furthermore,
$\Phi_\text{tot}$ is related to the physical flux by $\Phi_\text{tot} =
T_\text{obs} A_\text{eff} \phi_\text{tot}$, where $T_\text{obs}$ and
$A_\text{eff}$ denote observation time and instrument effective area,
respectively.
%We define for convenience the function (which resembles a photon histogram in
%the unbinned limit)
%\begin{equation}
%  \mathcal{C}(E) = \sum_{i=1}^{n_\text{ev}} \delta(E-E_i)\;,
%\end{equation}
%which has the property that, when averaged over many measurements,
%$\langle\mathcal{C}(E)\rangle = \Phi_\text{tot}(E)$.  We can then write for the
%log-likelihood
%\begin{equation}
%  -\ln\mathcal{L}(\vect \theta|\mathcal{D})
%  = \int_{E_\text{min}}^{E_\text{max}} dE\left(\Phi_{\text{tot}}(E)-\mathcal{C}(E)
%  \ln\Phi_{\text{tot}}(E|\vect\theta)\right)\;.
%\end{equation}
%
Furthermore, we assume that the model is linear,
\begin{equation}
  \Phi_\text{tot}(E|\vect\theta) = \sum_{k=1}^{n_\mathrm{comp}} \theta_k
  \Phi_k(E)\;.
\end{equation}

\begin{figure}[t!]
  \includegraphics[width=\columnwidth]{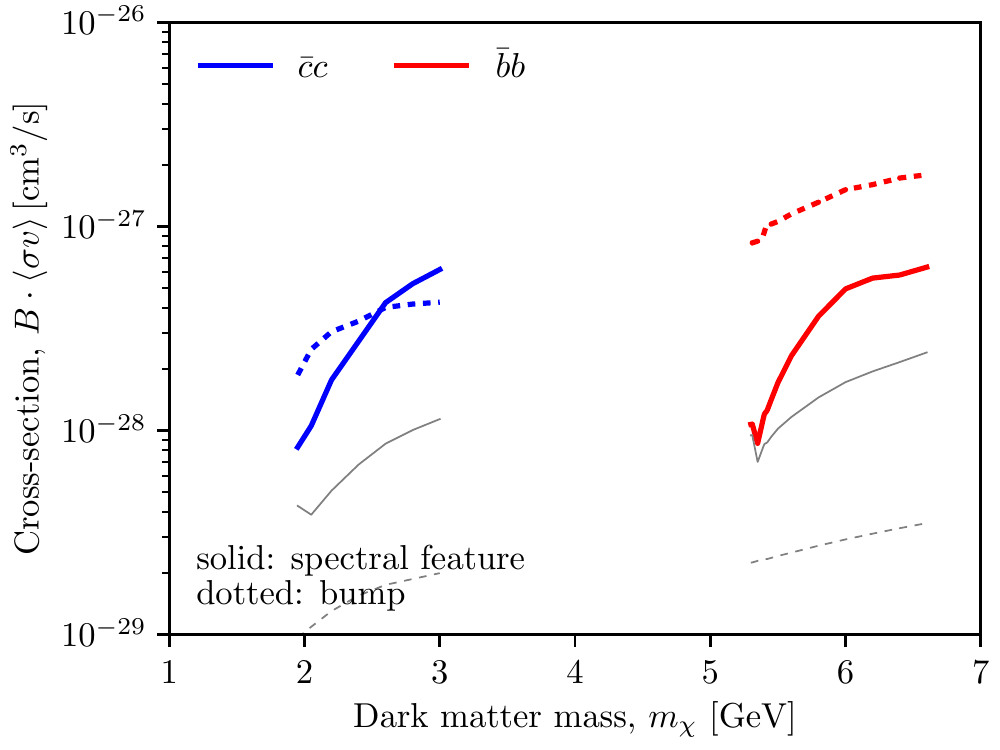}
  \caption{Same as Fig.~3 %\ref{fig:UL} 
  but for an idealized future experiment with 1\% energy resolution 
  and an effective area 10 times that of ComPair.}
  \label{fig:UL_toy}
\end{figure}

The Fisher information matrix is defined as the expected value of the second
moment of the score, i.e.~the gradient of the log-likelihood, averaged over multiple 
identical experiments.  In the present example, one can show that the Fisher information 
matrix is given by
\begin{equation}
%  \frac{\partial^2(-\ln\mathcal{L})}{\partial \theta_{k_1}\partial\theta_{k_2}}
  \mathcal{I}_{ij}(\vect\theta) = \int dE\,
  \frac{\Phi_{i}(E)\Phi_{j}(E)}{\Phi_{\text{tot}}(E|\vect\theta)}\;.
\end{equation}
This matrix
%is commonly referred to as Fisher information matrix, and 
is equivalent to Eq.~(2), %\eqref{eqn:fish}, 
where we used that further external constraints on the variance of the model parameters
can be implemented by adding the inverse of the variance to the corresponding
diagonal of the matrix.

The inverse of the Fisher matrix provides an approximation to the covariance matrix of the 
parameters of interest, which is used to derive the constraints and projections in this article. 
The diagonal entries of $\mathcal{I}^{-1}$ hence provide
estimates for the variance of the corresponding parameters, and their square root
an estimate for the variance of the corresponding number of standard deviation.
A one-sided
$95\%$\,CL upper limit corresponds to 1.65 standard deviations, because
integrating a standard normal Gaussian distribution from $-\infty$ to $1.65$
yields 0.95,
and hence in our case $\sqrt{(\mathcal{I}^{-1})_{55}}=1.65$. A $5\sigma$ 
{\it detection}, on the other hand, corresponds to 5 standard deviations and would 
therefore require a flux approximately three times larger than the upper
limits presented in Figs.~3, %\ref{fig:UL}, 
\ref{fig:UL_AG} and~\ref{fig:UL_toy}.
More details will be presented in Ref.~\cite{Fisher}.  We tested our results
with a conventional profile likelhood analysis \cite{Cowan:2010js}, and find identical results.

\bibliography{MeVbiblio}

\end{document}